# Photodiode Response in a CH$_3$NH$_3$PbI$_3$/CH$_3$NH$_3$SnI$_3$ Heterojunction

M. Spina, L. Mihály*, K. Holczer**, B. Náfrádi, A. Pisoni, L. Forró, E. Horváth

Laboratory of Physics of Complex Matter, Ecole Polytechnique Fédérale de Lausanne,

CH-1015 Lausanne

* Department of Physics and Astronomy, Stony Brook University, Stony Brook, NY 11790, USA

** Department of Physics and Astronomy, UCLA, Los Angeles, CA 90095-1547, USA

Since the discovery of its photovoltaic properties [1] organometallic salt CH$_3$NH$_3$PbI$_3$ became the subject of vivid interest. The material exhibits high light conversion efficiency [2,3], it lases in red color [4], and it can serve as the basis for light emitting diodes [5] and photodetectors [6,7]. Here we report another surprising feature of this material family, the photo-tunability of the diode response of a heterojunction made of CH$_3$NH$_3$PbI$_3$ [8] and its close relative, CH$_3$NH$_3$SnI$_3$ [9]. In the dark state the device behaves as a diode, with the Sn homologue acting as the "p" side. The junction is extremely sensitive to illumination. A complete reversal of the diode polarity, the first observation of its kind, is seen when the junction is exposed to red laser light of 25mW/cm$^2$ or larger power density. This finding opens up the possibility for a novel class of opto-electronic devices.

CH$_3$NH$_3$PbI$_3$ belongs to the family of methyl ammonium metal halide compounds, with the general composition of MA-MX$_3$, where MA stands for methylammonium, M is a metal ion (e.g. Pb$^{2+}$, Sn$^{2+}$) and X represents Cl$^-$, Br$^-$, I$^-$ or F$^-$. The [MX$_6$]$^{4-}$ octahedra are the building blocks of a perovskite structure. For the current study, large single crystals (MAPbI$_3$: 5-8mm; MASnI$_3$: 1-2mm in size), expressing cube-like or needle-shape crystal habit have been prepared by precipitation from a concentrated aqueous solution as described in Ref [10]. The resistivity and the I-V characteristics of the materials have been studied in the dark state and also under illumination by white light and as a function of the temperature [11]. Representative I-V curves illustrating the photo-response are shown in the Supplemental Material, Section 1.

Junctions were prepared by touching a ~1 mm-size single crystal of MAPbI$_3$ to the surface of a freshly prepared MASnI$_3$ platelet. Electrical contacts were made by glued gold wires or tungsten pins pressed to the top of the MAPbI$_3$ crystal and to the MASnI$_3$ crystal close to the junction. A representative device is shown in Figure 1a. Owing to the presence of Iodine vapor at the material's surface, the electrical contacts were susceptible to degradation with time, but the well-formed MAPbI$_3$ - MASnI$_3$ junctions were stable. We tested hundreds of such junctions and a subset of the devices – whose resistance was dominated by MAPbI$_3$ - MASnI$_3$ interface - exhibited the behavior described below.

The diode-like rectifying behavior of the device is evident in the I-V characteristic (Figure 1b). In the forward bias regime (V<0) the behavior looks nearly ohmic with a slope corresponding to $3 \times 10^9$ Ω. Note that the resistance of the bulk MAPbI$_3$ component of the device falls in the same order of magnitude and it is very likely that in the forward bias regime the current-limiting factor is the ohmic resistance of the crystal. In the reverse bias direction we do not see a complete saturation of the current. This feature is contrary to the behavior of a p-n junction or a heterojunction, but consistent with a metal-semiconductor junction, where the reduction of the Schottky barrier by the image charges results in a voltage-dependent reverse current. The slope of the I-V curve corresponds to a resistance of $4 \times 10^{10}$ Ω.

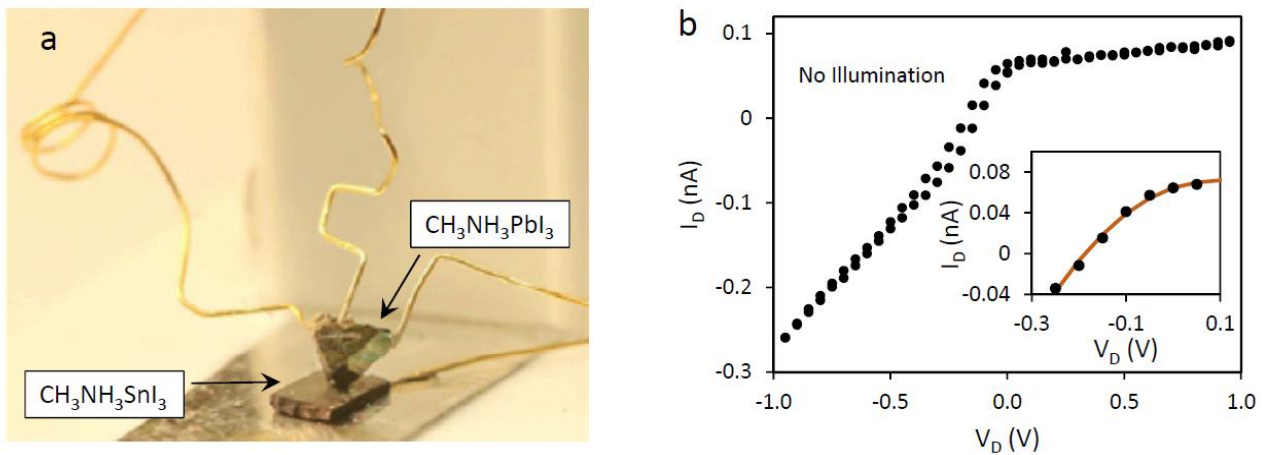

Figure 1. (a) Photograph of a photodiode similar to the ones used in this study. The MAPbI$_3$ bulk single crystal is mechanically compressed to a platelet of MASnI$_3$. Here the sample is contacted by gold wires, but most of the measurements were done in a two-probe configuration with tungsten pins pressed to the crystals. (b) Current-voltage characteristic of the diode in the dark state. The inset shows the low-voltage behavior (dots) and a fit to a simple thermionic model (line). In the forward-biased region the Sn homologue is the positive terminal.

At zero current (open circuit) the device generates a voltage of $V_{OC}$=0.17V; at zero bias there is a short circuit current of $I_{SC}$ = 64pA. In these freshly made samples the surfaces are still chemically active. It is likely that there is a transfer of ions between the two sides of the junction, resulting in an electrochemical voltage generation. At zero bias this "battery voltage" drives the short circuit current.

In the low-bias regime, where the junction resistance is high, we may neglect the ohmic resistance of the bulk $MAPbI_3$ and we may model the junction with a simplified thermionic emission model [12]. The ideality factor ($n$) and the barrier height ($U$) of the junction has been evaluated [Supplement 2]. The ideality factor, $n \geq 1$, gives the degree of the deviation of the diode characteristics from that of an ideal diode ($n=1$). The red curve in the inset of Fig. 1b represents a fit to the data yielding an ideality factor of $n=9.7$. The large value of $n$ indicates that the junction has inhomogeneities, probably related to the imperfect interface between the two materials. The same analysis yields a barrier height of $U = 0.92$eV. These values are reasonable, although the voltage range where the thermionic model works is rather limited.

The main finding reported here is the striking change in the behavior of the junction under illumination. Figure 2 shows the I-V curves of the device when the lead homologue is irradiated with red laser light ($\lambda=633$nm, spot size 4mm$^2$). At low light intensities (i.e. <5$\mu$Wcm$^{-2}$) reverse bias current evolves so that the junction becomes almost fully ohmic. As the voltage is scanned up and down, there is a hysteresis loop in the reverse biased region. With increasing light intensities (Figure 2b), the overall conductivity of the device increases dramatically, and at the same time there is a complete reversal of the forward and reverse bias behaviors. The ideality factor extrapolated for the device at different irradiation intensities varies between $n=13$ and $n=25$. The barrier height decreases approximately proportional with the light intensity reaching the value of 0.79eV for 250mWcm$^{-2}$.

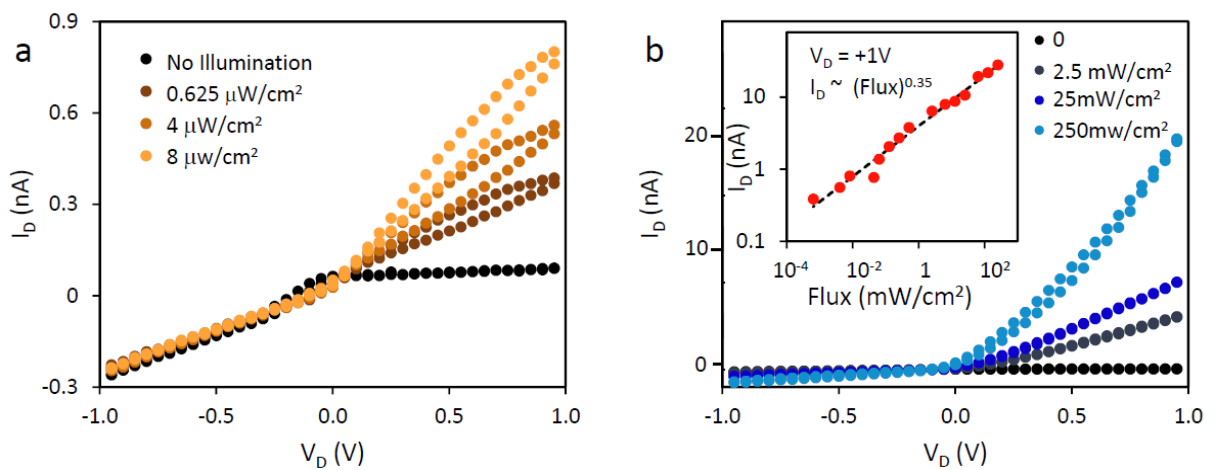

Figure 2. (a) Output characteristics of the photodiode at low light intensities (<10$\mu$Wcm$^{-2}$). Under illumination the lead compound starts conducting and the I-V curve has an almost Ohmic shape with a hysteretic behavior for positive voltage. (b) Output characteristics of the photodiode for higher light intensities. The inset shows the current at the bias voltage of +1V as the function of the light flux. The dashed line is a power-law fit.

We evaluated the photo response of the device as the function of incident light flux, as shown in the inset of Fig. 2b. For positive bias of +1V the current follows a power law dependence with an exponent of 0.35, for incident flux ranging from 0.625μW/cm$^2$ to 250mW/cm$^2$. There is no sign of saturation. The steady increase of current for nearly 6 orders of magnitude of incident flux may be an extremely useful feature for light-sensing applications.

It is clear from the shape of the I-V curves in Fig. 2 that the photo-response of this device cannot be explained by the usual photoelectric effect, when electron-hole pairs created by the incident light carry a current through the device. Instead, we will use a model we proposed in an earlier study to explain the dramatic photoconductivity observed on the lead homologue [11]. The key element of the model is the introduction of a collection of narrow impurity states close to the conduction band. Charge carriers in this band are localized, but - in contrast to the usual doped semiconductors - in the dark state these states are unfilled and they can be attributed to charge neutral vacancies [13]. We have also postulated a hierarchy of relaxation rates (in particular a long relaxation rate from the impurity states to the valence band) that results in a "bottleneck" on the charge carrier's recombination path from the conduction band to the valence band.

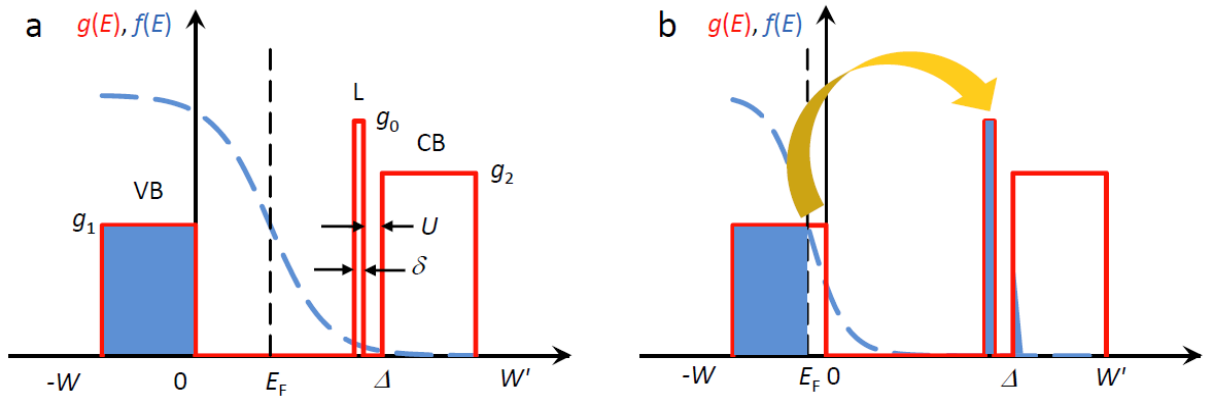

Figure 3. (a) Schematic density of states, $g(E)$, of MAPbI$_3$. The filled valence band (VB), the empty localized states (L) and the conduction band (CB) are indicated. The dashed line is the Fermi function $f(E)$, centered on the Fermi energy $E_F$. For a better view the width of the Fermi function exaggerated. (b) Electrons excited by light get trapped in the localized states, causing a shift of the Fermi energy, discussed in the text.

Figure 3.a illustrates the schematic band structure and the electron occupation in the ground state. We assume a flat form for the density of states $g(E)$, but the main features of the results will be the same for a more realistic functions. The Fermi energy is determined by the condition $N = V \int_{-W}^{W'} g(E) f(E) \, dE$, where $N$ is the number of electrons and $-W$ and $W'$ represents the bottom of the valence band and the top of the conduction band, respectively. Assuming that temperature is less than the bandwidths and the bandgap ($kT<<\Delta, W$), and that the width of the localized band is much less than its separation from the conduction band ($U<<\delta$), the Fermi energy is approximately determined by

$$0 = \int_{-W}^{W'} g(E) f(E) \, dE - \int_{-W}^{0} g(E) \, dE = -g_1 kT e^{-\frac{E_F}{kT}} + g_2 kT e^{-\frac{\Delta - E_F}{kT}} + U g_0 e^{-\frac{\Delta - \delta - E_F}{kT}}$$

For $g_1=g_2$ and $g_0=0$ this yields the well-known result of $E_F=\Delta/2$ (independent of temperature). The unoccupied localized states push the Fermi level to a (temperature dependent) lower value. For example, if $g_1 kT = g_2 kT = U g_0$, then $E_F=\Delta/2 - kT/2 \ln(1+e^{\delta/kT})$.

Photo-excitation changes this picture drastically, due to the bottleneck effect described above. There will be a new dynamic equilibrium governed by the rates of excitation and the relaxation of the charge carriers, and the localized states will acquire a significant occupation. To illustrate the trends caused by this effect, we take the extreme case, when photo-excitation nearly fills the localized states, as shown in Figure 3. The electrons trapped in the localized states leave behind an equal number of holes in the valence band. If thermal relaxation rates in the valence band are sufficiently high, the holes will thermalize around a new Fermi energy derived from $N - \delta N = V \int_{-W}^{W'} g(E) f(E) \, dE$, where $\delta N$ is the number of trapped electrons. The result is illustrated in Figure 3.

Although here we assumed a narrow band of localized states, similar conclusions are achieved if the localized states are not separated by a gap $\delta$ from the conductions band, but instead the conduction band itself has a tail of localized states. Such states, separated from the delocalized states by the "mobility edge", has been widely discussed for disordered systems [14] and also for the materials used in the current study [15,16]

The dramatic change of the Fermi energy in the photo-excited sample explains the reversal of the junction observed in the experiment, as illustrated in Figure 4. The Fig. 4a shows the flat-band energy diagram of the two materials, with an energy scale relative to the vacuum energy, based on band structure calculations [17, 18]. Although structurally similar, MAPbI$_3$ and MASnI$_3$ have different electronic properties [19,20], as reflected in the position of the Fermi energy. MAPbI$_3$ is an ambipolar semiconductor with a band gap of 1.5eV, and the Fermi energy is approximately in the middle of the band gap. The resistivity of our MASnI$_3$ crystals vary between thermally activated (with an activation energy of 250meV) and metallic, with a large range of room temperature resistances ($10^3$ to $10^{-3}$ Ωcm). Earlier studies demonstrated that this material can be easily doped to the metallic temperature dependence [21]. The variation in the properties of the MASnI$_3$ crystals can be attributed to the fact that the preparation involves the reduction of Sn$^{IV}$ to Sn$^{II}$ and a partial reduction results in a doped material [16]. For the current studies we selected a low resistance (highly doped) MASnI$_3$ crystal, and for those crystals the Fermi energy is near to the top of the valence band.

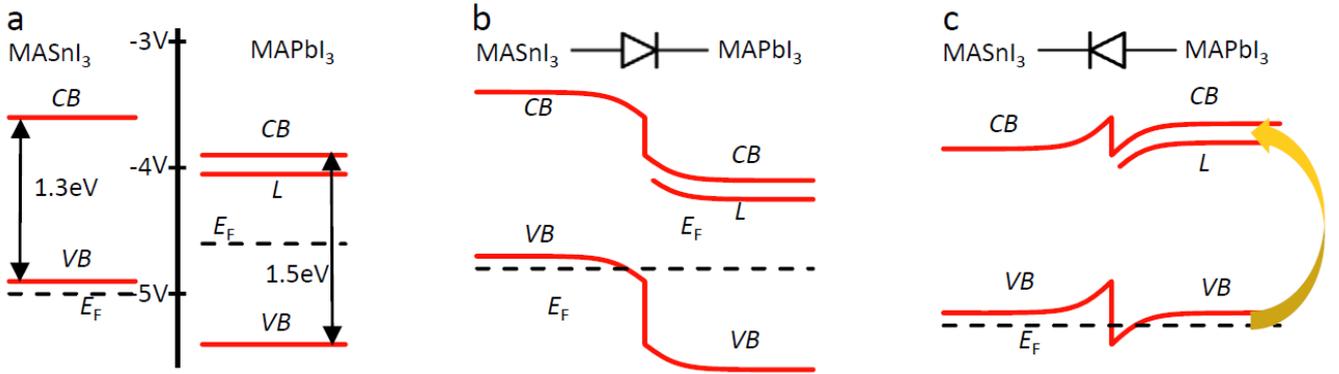

Figure 4. (a) Flat-band diagram of the two materials, based on band structure calculations [17]. The energies are relative to the vacuum energy. (b) Heterojunction formation in the dark state. (c) Electrons excited by light get trapped in the localized states, causing a shift of the effective Fermi energy and a reversal of the junction.

In the dark state the matching of the Fermi levels of MASnI$_3$ and MAPbI$_3$ yields a junction where the MASnI$_3$ is p-type relative to the MAPbI$_3$ (Fig. 4b). When the photo-excitation shifts the Fermi level of the MAPbI$_3$, we have a p-P heterojunction, where the roles are switched and MAPbI$_3$ becomes the positive terminal of the diode. (Fig. 4c).

The electrons trapped in the localized states cannot participate in the diffusion of charge carriers that typically happens when the heterojunction is formed. As a secondary effect, trapped localized charges in the interface region may also be responsible for the hysteresis seen in the low-bias I-V curves, shown in Figure 2a. The regular capacitive effects typical of a junction with a space-charge region result in a relatively small capacitance, and this is not sufficient to explain the observed hysteresis at the low scanning rate used in our study. The response of the trapped charges can lead to a long-term memory in the junction, yielding to the observed behavior.

In summary, we report a large photo-response in a $CH_3NH_3PbI_3/CH_3NH_3SnI_3$ heterojunctions that steadily increasing with the incident flux for over more than 5 orders of magnitude of flux range. The behavior of the junction is different from the traditional photodiodes. In particular, the current in the junction is not carried by photo-excited electron hole pairs. Instead, under illumination we see a dramatic reversal of the diode's polarity. We propose an explanation based on a model where the incident light effectively induces an insulator to metal transition in the $CH_3NH_3PbI_3$ side of the junction. This is the very same model that has been used earlier to explain the photoresponse of the lead homologue [11], We believe the unique properties described by the model are essential for this material's success in photo-voltaic applications.

To our knowledge this is the first observation of such a dramatic reversal of polarity in a diode under illumination. Just like the first diodes (made of semiconductors and metal whiskers) or the first point-contact transistor made by Bardeen, Brattain and Shockley, our device is rudimentary, but it has the potential for novel applications, including the construction of extremely sensitive light sensors and other optoelectronic devices.

Acknowledgments. The work was supported by the ERC Advanced Grant Picoprop of L.F. The hospitality of EPFL is gratefully acknowledged by L.M. and K.H.